# GAMER-MRIL identifies Disability-Related Brain Changes in Multiple Sclerosis

Po-Jui Lu[1,2,3], Benjamin Odry[4], Muhamed Barakovic[1,2,3], Matthias Weigel[1,2,3,5], Robin Sandkühler[6], Reza Rahmanzadeh[7], Xinjie Chen[1,2,3], Mario Ocampo-Pineda[1,2,3], Jens Kuhle[2,3], Ludwig Kappos[2,3], Philippe Cattin[6], Cristina Granziera[1,2,3*]

[1] *Translational Imaging in Neurology (ThINk) Basel, Department of Biomedical Engineering, University Hospital Basel and University of Basel, Basel, Switzerland*

[2] *Department of Neurology, University Hospital Basel, Switzerland*

[3] *Research Center for Clinical Neuroimmunology and Neuroscience Basel (RC2NB), University Hospital Basel and University of Basel, Basel, Switzerland*

[4] *AI for Clinical Analytics, Covera Health, New York, NY, USA*

[5] *Division of Radiological Physics, Department of Radiology, University Hospital Basel and University of Basel, Basel, Switzerland*

[6] *Center for medical Image Analysis & Navigation, Department of Biomedical Engineering, University of Basel, Allschwil, Switzerland*

[7] *University Institute of Diagnostic and Interventional Neuroradiology, Inselspital, Bern University Hospital, University of Bern, Bern, Switzerland*

*Abstract*—*Objective:* Identifying disability-related brain changes is important for multiple sclerosis (MS) patients. Currently, there is no clear understanding about which pathological features drive disability in single MS patients. In this work, we propose a novel comprehensive approach, GAMER-MRIL, leveraging whole-brain quantitative MRI (qMRI), convolutional neural network (CNN), and an interpretability method to classify MS patients with severe disability and investigate relevant pathological brain changes. *Methods:* One-hundred-sixty-six MS patients underwent 3T MRI acquisitions and were divided into cross-validation and test datasets. qMRI informative of microstructural brain properties was reconstructed, including quantitative T1 (qT1), myelin water fraction (MWF), and neurite density index (NDI). To fully utilize the qMRI, GAMER-MRIL extended a gated-attention-based CNN (GAMER-MRI), which was developed to select patch-based qMRI important for a given task/question, to the whole-brain image. To find out disability-related regions, GAMER-MRIL modified a structure-aware interpretability method, Layer-wise Relevance Propagation (LRP), to incorporate qMRI. *Results:* The test performance was AUC=0.885. qT1 was the most sensitive measure related to disability, followed by NDI. The proposed LRP approach obtained more specifically relevant regions than other interpretability methods, including the saliency map, the integrated gradients, and the original LRP. The relevant regions included the corticospinal tract, where average qT1 and NDI significantly correlated with patients' disability scores ($\rho$=-0.37 and 0.44). *Conclusion:* These results demonstrated that GAMER-MRIL can classify patients with severe disability using qMRI and subsequently identify brain regions potentially important to the integrity of the mobility function. *Significance:* GAMER-MRIL holds promise for developing biomarkers and increasing clinicians' trust in NN.

## I. INTRODUCTION

MULTIPLE Sclerosis (MS) is a chronic inflammatory disease of the central nervous system. The representative microstructural characteristics include multifocal inflammatory infiltration, demyelination, remyelination, and axonal loss leading to the accumulation of disability in MS patients [1]. In the clinic, disability in MS patients is assessed by the Expanded Disability Status Scale (EDSS), a nonlinear representation of clinical disability ranging from 0 to 10. Starting from 5, MS patients are considered to have a severe disability. So far, however, it has been quite challenging to identify which pathological changes drive the accumulation of disability in MS patients. Often summary measures are used (e.g., brain/spinal cord atrophy, number of lesions or specific lesion types, etc.) but those measures do not permit us to understand the mechanisms underlying the accumulation of deficits in specific patients with MS.

Quantitative magnetic resonance imaging (qMRI) provides biophysical measures of microstructural properties of the central nervous system: myelin properties can be quantified using surrogate imaging markers of myelin integrity, such as myelin water fraction (MWF); axonal characteristics might be studied through measures derived by modeling the diffusion signal, such as neurite density index (NDI), and global information about the microstructural environment of the tissue might be gained through quantitative T1 relaxometry (qT1) [2]. These qMRI measures can be used to gain insight into the pathological mechanisms driving disease progression and clinical worsening in MR.

Deep Neural Networks (DNNs) have been used in the neuroimaging field and in specific cases, they have also been applied to qMRI data. In this context, we have recently developed a method – GAMER MRI – which exploits attention weights (AWs) in a CNN based on gated attention mechanism to select which qMRI measures were most important for MS and stroke lesions classification [3] and which enhances the correlation with clinical and biological measures [4]. However, the deep layer structure and nonlinear activations represent a limitation on the interpretability of the decision process of DNNs [5]. This becomes a major challenge for medical applications where evidence-based clinical decisions are needed. To tackle this issue, several methods have been developed to improve interpretability, including saliency [6], integrated gradients [7], and layer-wise relevance propagation (LRP) [8]. LRP has been shown to be more effective than other methods and illustrated the disease-specific evidence [5], [9]–[12]. It leverages the DNN layer structure to generate a map, which is based on designed rules and the logits f(x) from the classifier, showing the relevance of individual pixels to the given task [13], [14].

Nevertheless, current LRP rules do not fully exploit the information provided by gated attention mechanism and the quantitative nature of qMRI. Relevance values only propagate through part of the element-wise multiplication structure in the long short-term memory network (LSTM) [13]. This structure is also utilized in the gated attention mechanism. Gated attention mechanism focuses on the most relevant features and this partial propagation can reduce the explanation power of the obtained relevance values for attention-based CNN. In this work, we investigated whether (i) an LRP relevance map based on AWs is more relevant than a map based on the logits of a DNN, and whether (ii) the LRP relevance maps based on AWs could be linearly combined with qMR images so that it is possible to identify areas in qMR images that are related to clinical disability, as measured by EDSS. Following the results, we proposed GAMER-MRIL incorporating the improved GAMER-MRI and the new LRP-based approach for classification of patients with severe disability using whole-brain microstructural qMR images and for identification of relevant areas. A preliminary version of this work has been reported in ECTRIMS 2022 and ISMRM 2023 [15], [16].

## II. METHODS

### A. MRI data

One-hundred-sixty-six MS patients (100 RRMS and 66 PMS, 99 females and 67 males, age range=45.9±14.3 years, median EDSS=2.5) were enrolled in the study, which was approved by the ethics committee. Written consent was obtained prior to the MRI acquisition. Forty out of the 166 patients had EDSS ≥5 (severe disability group) and 126/166 had a mild disability (<5). Sixty-six patients had a two-year follow-up acquisition. Patients underwent a multi-parametric protocol on a 3T Siemens Healthineers MAGNETOM Prisma MRI system. The volumetric protocol included 3D Magnetization-Prepared 2 RApid Gradient Echoes (MP2RAGE, $1\,mm$ x $1\,mm$ x $1\,mm$ and an image size of 176x240x256 voxels) [17], 3D turbo spin echo (SPACE) based FLuid Attenuated Inversion Recovery (FLAIR) ( $1\,mm$ x $1\,mm$ x $1\,mm$, 176x240x256 voxels), 3D Fast Acquisition with Spiral Trajectory and T2prep sequence (FAST-T2, $1.25\,mm$ x $1.25\,mm$ x $5\,mm$, 256x256x32 voxels) [18], and multi-shell Diffusion-Weighted Imaging (multi-shell DWI, $1.8\,mm$ x $1.8\,mm$ x $1.8\,mm$, an image size of 142x142x80 voxels, 12 images of a b-value = 0, 137 images of b-values = 700, 1000, 2000 and 3000 $s/mm^2$) with contiguous 2D slices.

From multi-parametric MRIs, qMR images were further reconstructed. The qT1 was reconstructed from MP2RAGE as in [17]. The MWF map was reconstructed from FAST-T2 as in [18]. The neurite density index (NDI) from the neurite orientation dispersion and density imaging model (NODDI) [19] was reconstructed from multi-shell DWI as in [20]. By FSL and FreeSurfer [21], [22], co-registration of qMR images and removal of non-brain tissue were performed. qT1 feasible values fall between 0 and 2500 $ms$ excluding ventricles [23], [24]. MWF feasible values within the brain are at most 30% [25]. NDI is between zero and one. qMRI measures were then normalized between zero and one. Normalized qMR images were transformed to the coordinates of NDI. NDI was chosen as the reference coordinate for the trade-off between the lower degree of slice interpolation for MWF and the retainment of fine resolution of qT1. The appearance of the cerebellum in NDI maps was coarse with limited information and therefore it was removed to improve the performance of the subsequent training. The matrix size of each qMR image was (96, 96, 112) after the removal of empty space.

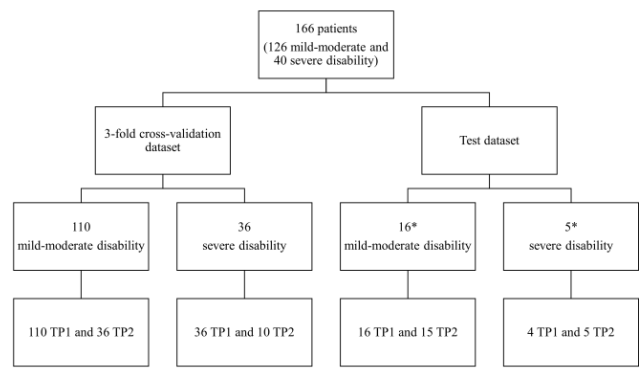

Fig. 1. The data split. TP1 refers to a patient's baseline acquisition and TP2 refers to the follow-up acquisition. *: A patient's EDSS score became 6 at the follow-up acquisition.

The dataset was divided into a test dataset and a dataset for cross-validation as in Fig. 1. In this case, the dataset for cross-validation was used for the optimization of the hyperparameters and the model selection. The test dataset was used to estimate the model performance on unforeseen data. In light of the limited number of patients in the severe group, stratified 3-fold cross-validation was used. A patient's baseline and follow-up acquisitions were considered two samples in the same fold and the same dataset. As a result, in each round of cross-validation, two folds formed the training dataset and the remaining fold was the validation dataset.

*B. GAMER-MRI*

The core idea of multi-path GAMER-MRI was to use the gated attention mechanism and a parallel encoding structure to generate AWs as proxies of relative importance among multi-contrast MR images [3]. The variant gated attention mechanism is formulated as follows:

$$\boldsymbol{n} = \sum_{l=1}^{L} a_l \boldsymbol{m}_l = \sum_{l=1}^{L} a_l q_l(x_l), \quad (1)$$

where $\boldsymbol{n}$ is the combined representation for the classifier, $\boldsymbol{m}_l \in R^{(M \times 1)}$ is the hidden representation of the $l^{th}$ contrast, $q_l(x)$ is the corresponding encoding function and $a_l$ is the AW of the $l^{th}$ contrast. The AW is based on the outputs from the signal and gate branches, $s$ and $g$, respectively, in Fig. 2 and is given by:

$$a_l = softmax(\boldsymbol{w}^T(\boldsymbol{s} \odot \boldsymbol{g})), \quad (2)$$

$$\boldsymbol{s} = tanh(\boldsymbol{U}\boldsymbol{m}_l); \boldsymbol{g} = sigm(\boldsymbol{V}\boldsymbol{m}_l), \quad (3)$$

where $\boldsymbol{U}$ and $\boldsymbol{V} \in \boldsymbol{R}^{(K \times M)}$ are corresponding weights of the fully connected layers (FCs) in the signal and gate branches, $tanh$ stands for the nonlinear hyperbolic tangent function, $sigm$ is the nonlinear sigmoid function, $\odot$ is the element-wise multiplication operator, $w \in R^{(1 \times K)}$ contains the weights of the last FC.

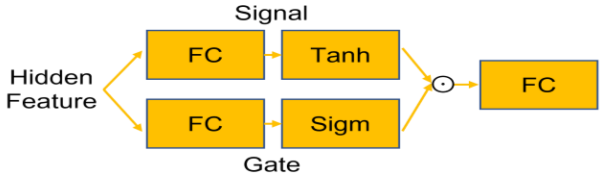

Fig. 2. The gated attention mechanism. Each FC stands for a fully connected layer. $\odot$ represents an element-wise multiplication.

*C. Network in GAMER-MRIL*

The multi-path GAMER-MRI included three main compartments, including the parallel feature extraction, gated attention mechanism, and a classifier, as depicted in Fig. 3.

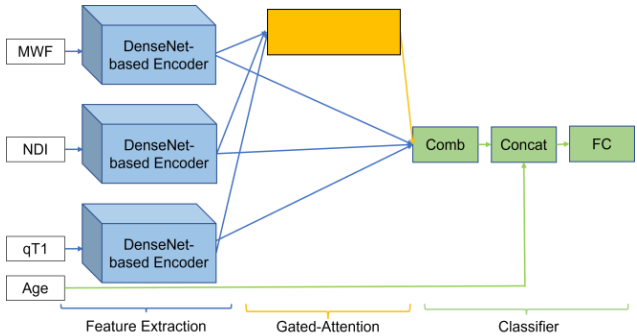

Fig. 3. The network structure. Comb is linear combination of attention weights and hidden features from the encoders. Concat is the concatenation of the age information. FC is the fully connected layer.

Different from the original GAMER-MRI [3], [4], we used the DenseNet [26] architecture for the feature extracting compartment in GAMER-MRI. The average number of voxels in a volumetric MR image is on the order of $10^6$. DenseNet concatenates all feature maps generated within the same dense block to reuse feature maps and facilitate feature propagation. This reduces the number of trainable parameters substantially and alleviates the vanishing gradient issue. The feature extraction consisted of a convolutional block, four dense blocks, and three transition layers. The convolutional block was composed of an initial convolutional layer (Conv) of 16 filters with a kernel size of 3x3x3, a batch normalization layer (BN), rectified linear units (ReLU) and a 3D max pooling layer with a kernel size of 3, a stride of 2 and a padding of 1. Each dense block contained two dense layers, each of which was a collection of BN, ReLU, and Conv with a kernel size of 1x1x1 and replication padding, and BN, ReLU, and Conv with a kernel size of 3x3x3. The transition layer was BN, ReLU, Conv (1x1x1), and average pooling with a kernel size of 2x2x2 and a stride of 2. The growth rate of the number of feature maps was four. The fourth dense block was followed by a BN, ReLU, and a FC of 32 neurons to form a hidden feature vector.

In gated attention mechanism, the FCs in the signal layer and the gate layer had 16 neurons. The combined hidden feature vector, which was formed by the AWs and the hidden feature vectors from all paths, was concatenated by patients' age information. Patients' ages were divided by 100 prior to concatenation. The incorporation of age information accounted for the age effect [27].

The classifier was an FC of a neuron outputting f(x) that was subsequently transformed to classification probability by $sigm$.

*D. Training strategy*

For the loss function, the binary cross-entropy was used. In consideration of the heterogeneity within each group and across groups measured by EDSS, the training loss of each sample was weighted by the closeness between EDSS=5 and patients' EDSS (weight=$2 - (|EDSS - 5|)/5$). The mini-batch size was 70 for training. The weighted sampler was used to account for the class imbalance during training, and the optimizer was the Adam optimizer with decoupled weight decay (AdamW) with a learning rate=5e-5 and the default weight decay=1e-2 [28]. The evaluation metric was the area under the receiver operating characteristic curve (AUC). To alleviate overfitting, data augmentation included random flipping, random 90-degree rotation, random Gaussian noise of zero mean and standard deviation equal to 0.2, and random affine transformation with maximum rotation ±30 degrees and ±10% scaling. We implemented our method in PyTorch v1.7 and used two A100 GPUs (Nvidia, Santa Clara, CA, USA) for training.

*E. Layer-wise Relevance Propagation*

LRP is a post-hoc explaining technique for neural networks and is based on Deep Taylor Decomposition (DTD) [29]. The main rationale is to redistribute the prediction backward through the layer structure of the DNN to the input data based on the percentage of contribution from the individual neuron in the forward pass using defined redistribution rules. It is formulated as follows:

$$R_x = D_j\left(D_{j+1}\left(\ldots D_c(f(x))\right)\right), \quad (4)$$

$$\sum_{v \in x} R_v = f(x), \quad (5)$$

where $R_x$ is the relevance map of the input image $x$, $R_v$ is the relevance of the voxel $v$, $D_c$ is the redistribution rule for the classifier and $D_j$ is the rule for the $j^{th}$ intermediate layer. For (5) to be valid, the bias parameters in the DNN aren't considered in the LRP backward passes to get the relevance map [5]. For the rest of the paper, all the rules do not consider the bias parameters. The redistribution rules depend on the kinds of layers and are thus versatile. The common rules include the ε-rule, the αβ-rule, the $z^\beta$-rule and others [13], [14]. The ε-rule incorporates a small ε value to avoid zero division and was designed for the FC. The αβ-rule was designed for the Conv separating the positive and negative contributions. The $z^\beta$-rule was designed for the layer taking the input image to consider the upper and lower bounds of the image. In [14], they recommended different rules for layers in different positions in the hierarchical structure and the kinds of layers, so we applied an orderly mixture of rules in this study.

### F. New LRP approach in GAMER-MRIL

The approach consisted of three strategies. (i) To account for the element-wise multiplication $s \odot g$ in (2), we proposed a new rule. In [13], they proposed a LRP-all rule for the element-wise multiplication of the cell input and the input gate within the long short-term memory network (LSTM) and showed that it was more suitable than the LRP-prop [30] and LRP-abs rules. The LRP-all rule only lets the cell input take all the relevance. However, for the gated attention mechanism, this rule neglects the nature of the structure that both $s$ and $g$ contribute towards the attention weights. The proposed rule is formulated as follows:

$$R_s = \frac{|s|}{|s| + |g|} R_a, \quad (6)$$

$$R_g = \frac{|g|}{|s| + |g|} R_a, \quad (7)$$

where $R_s$ is the relevance for the signal branch, $R_g$ is the relevance for the gate branch and $R_a$ is the relevance received by the attention weight. The rule is similar to LRP-abs, which in principle takes the absolute values of $Vm_l$ and $Um_l$, if it is applied to the gated attention mechanism. The scales of $Vm_l$ and $Um_l$ are not bounded, so the LRP-abs rule cannot properly reflect the individual contributions in the element-wise multiplication. For LRP to be applicable to the DenseNet, we followed suit in [31] to merge the collection of BN, ReLU, and Conv into equivalent Conv and the collection of BN and FC into equivalent FC during the LRP backward pass. Other layers in the NN used the same as in [14]. The ε in the ε-rule was 1e-8. The α and β were 1 and 0, respectively. Due to normalization of the qMRI measures, the upper and lower bound in $z^\beta$ were 1 and 0. For each qMR image, a relevance map was generated. (ii) Here, we proposed to start the LRP backward pass from AWs for the relevance map instead of the logits f(x). The AW as the proxy for the importance of the input image should focus on more relevant features. (iii) The parallel encoding structure of the network in GAMER-MRIL and the corresponding AWs can potentially allow a linearly combined map incorporating the joint information of the input qMR images.

$$Combined\ map = \sum_{x \in X} R_x * x \quad (8)$$

### G. Assessment

To assess if the proposed approach can reveal more important brain regions, a voxel value inverting experiment from the aspect of model performance and Spearman's correlation (ρ) with EDSS from the clinical aspect were conducted after the network was trained. The joint dataset from the test and the validation datasets were in use for both assessments to have a sufficiently large number of samples.

There were three conditions to be evaluated, including (i) whether the relevance map was based on AWs or f(x), (ii) if the proposed rule or the original LRP-all rule was used and (iii) whether the individual relevance maps or the combined map were/was considered. Table I lists the eight possible scenarios and four scenarios, where heatmaps based on the saliency and integrated gradient methods were compared. In the scenarios 5 and 6, where the LRP backward pass started from the f(x), the combined representation in (1) and the element-wise multiplication in the gated attention mechanism used the same rule. The relevance maps and the combined maps were binarized from high to low quantiles of positive relevance within the brain as in [10] to obtain quantile masks. In the voxel value inverting experiment, voxel values of the normalized qMR images were inverted, i.e., qMR = 1 - qMR, according to the quantile masks. If regions identified by a quantile mask are important, the AUC is affected by inverted voxel values in the regions and reduces. If a quantile mask in a scenario can reduce the AUC more than in the other scenarios, this means the mask identifies more important regions.

The scenario that achieved the largest drop in AUC was assessed for the correlation with patients' EDSS. The correlation was performed on the normalized qMR values, which were averaged within the quantile mask of the scenario.

**TABLE I**
**THE TWELVE SCENARIOS FOR COMPARISON**

| Scenarios | | | | |
|---|---|---|---|---|
| Start | AW | | $f(x)$ | |
| Rule | Proposal | LRP-all | Proposal | LRP-all |
| Individual Relevance Map | 1 | 2 | 5 | 6 |
| Combined Map | 3 | 4 | 7 | 8 |
| Rule | Saliency | Integrated Gradient | Saliency | Integrated Gradient |
| Individual Relevance Map | 9 | 10 | 11 | 12 |

The two-sided permutation test with 20,000 permutations was used for testing the strongest correlation. The statistics were performed in R. The quantile mask of the strongest correlation of each patient was further nonlinearly transformed by greedy in ITK-Snap [32] to the MNI152 template for exploring potential group effect areas on the heatmap.

## III. RESULTS

### A. Performance of patient classification

The test and averaged cross-validation results in AUC, accuracy, specificity, sensitivity, and the mean AWs of qMRIs are reported in Table II. The performance on cross-validation and test datasets indicated that the network learned a right representation to classify patients into severe and moderate/mild disability groups. This established the ground for further analyses related to LRP. As in [3], [4], the reported AWs were averaged across the correctly classified samples. qT1 was the most important, followed by NDI and MWF in terms of the AWs.

### B. Voxel Inverting Experiment to assess important regions

TABLE II
RESULTS OF THE TEST DATASET AND THE AVERAGE THREE-FOLD CROSS-VALIDATION

| Datasets | AUC | Accuracy | Specificity | Sensitivity |
|---|---|---|---|---|
| CV dataset | 0.864 | 0.809 | 0.839 | 0.718 |
| Test dataset | 0.885 | 0.854 | 0.844 | 0.889 |
|  | **MWF** | **NDI** | **qT1** | |
| Attention weights of the CV dataset | 0.188 | 0.309 | 0.503 | |
| Attention weights of the test dataset | 0.164 | 0.369 | 0.467 | |

The AUCs in all scenarios are given in Fig. 4. From the top $40^{th}$ quantile, i.e., the $60^{th}$ quantile, the scenario 3, where the quantile masks were defined by the proposed approach, had a lower AUC than the other scenarios.

### C. Correlation with EDSS

When qMR voxel values within the top $40^{th}$ quantile mask were averaged, ρ of qT1 and NDI with EDSS were the largest (Table III). Compared with ρ in the scenario 3, ρ of qT1 and NDI in the scenario 4 (i.e. using the original LRP rule, but starting the backward pass from the AWs and using the combined map) were smaller. Therefore, the permutation test was performed on the regions covered by the top $40^{th}$ quantile mask in the scenario 3, where the quantile mask was based on the proposed approach. The correlation was statistically significant (Table IV). The NDI images of two exemplary patients in the groups with mild and severe disability, as well as the top $40^{th}$ quantile masks, are shown in Fig. 5. In Fig. 5, the regions that are most related to EDSS overlap less with the MS lesions, which are near the posterior horn of the lateral ventricles. For the patient in the severe disability group, the relevant regions cover extensively the posterior limb of the internal capsule and the thalamus.

TABLE III
SPEARMAN'S CORRELATION COEFFICIENTS BY DIFFERENT QUANTILE MASKS.

| Top $N^{th}$ Quantile | qT1 | NDI |
|---|---|---|
| **Scenario 3** | | |
| 30 | -0.362 | 0.429 |
| 40 | -0.371 | 0.440 |
| 50 | -0.366 | 0.431 |
| **Scenario 4** | | |
| 30 | -0.269 | 0.337 |
| 40 | -0.281 | 0.348 |
| 50 | -0.277 | 0.345 |

TABLE IV
SPEARMAN'S CORRELATION COEFFICIENT IN THE SCENARIO 3 WITH A TOP 40TH QUANTILE MASK AND THE P-VALUES FROM THE PERMUTATION TEST.

|  | qT1 | NDI | MWF |
|---|---|---|---|
| 40 | -0.362 | 0.429 | 0.298 |
| P-value | 0.001 | 0.0001 | 0.01 |

### D. Heatmap on MNI152

Fig. 6 illustrates the heatmaps of the scenarios 3 and 4. Identified regions included the left thalamus, the left internal capsule, and part of the putamen. Scenario 3, which used the proposed approach, identified more regions, including the left caudate, a larger part of the right putamen, and the right internal capsule.

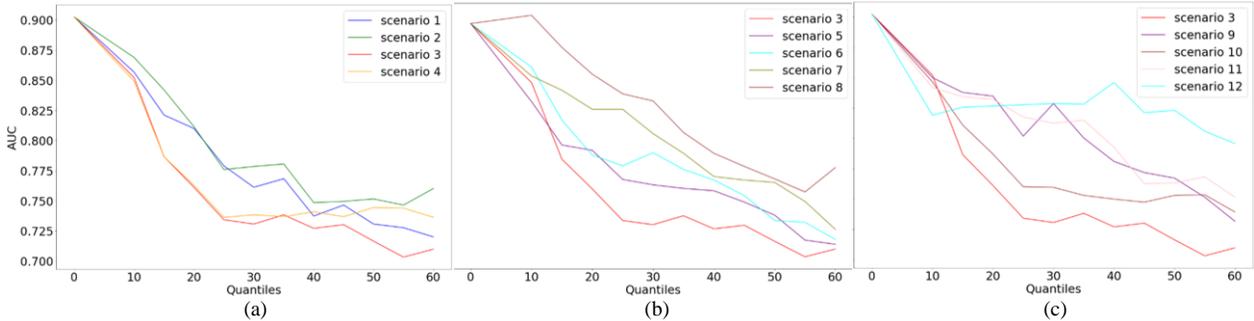

Fig. 4. The AUC results of the voxel inverting experiment of different top $N^{th}$ quantile masks. (a) The scenarios from 1 to 4 used the relevance maps based on the attention weights and the combinations of different LRP rules and the individual masks or combined mask.; (b) The scenarios from 5 to 8 and the scenario 3 used the relevance maps based on the output f(x) and the combinations of different rules and the individual masks or the combined mask.; (c) The scenarios from 9 to 12 and the scenario 3 used the heatmaps of the saliency method and the integrated gradient method based on the output f(x) and the attention weights.

## IV. Discussion

In this work, we demonstrated that GAMER-MRIL improved from GAMER-MRI for patch-based classification on lesions and classified MS patients with a severe or mild disability using whole-brain qMR images. Besides, we provided evidence that the proposed LRP approach in GAMER-MRIL – which consisted of three strategies, including the new rule for the gated attention mechanism, AW-based relevance and the combined relevance maps considering qMR images - could better identify brain regions whose alterations were most related to patients' disability.

The network in GAMER-MRIL showed that qT1 was the measure that best discriminated clinical severity in our cohort of MS patients, followed by NDI and MWF. This might be due to different reasons: (i) qT1 provides a more comprehensive view on tissue damage compared to NDI and MWF (i.e. global microstructural damage and iron accumulation vs axon and myelin-related damage) [2] or (ii) the original higher spatial resolution and the higher white-grey matter contrast of qT1 revealed more details. Furthermore, the new LRP approach in GAMER-MRIL identified the disability-related alterations. In the voxel value inverting experiment, the first four scenarios started the LRP from the AWs and compared combinations of the new rule, the original LRP-all rule, individual relevance maps as in the original LRP method and the proposed combined relevance map. The fifth to eighth scenarios started the LRP from the f(x) as in the original LRP method and compared the same combinations. The new LRP approach in GAMER-MRIL (scenario 3) resulted in the lowest AUC when the qMRI measures within the obtained relevant regions were inverted and thus the regions carried more information than the ones found by other conditions. This was also true when two other typical interpretability methods were compared (scenarios 9 to 12). This supports our hypothesis that AWs, used as a proxy for the importance of input qMR images, contain more relevant information than the output from the classifier.

The averaged qMRI measures within the regions identified by the LRP approach in GAMER-MRIL showed the best correlations. From a clinical perspective, this supports our hypothesis that those regions are not merely important to the DNN, but also meaningfully related to clinical measures of disability in MS patients. NDI provides specific information about axonal content [19], and qT1 quantifies the overall microstructural tissue environment [24]. Axonal damage and demyelination within lesions and in the normal-appearing tissue should increase the qT1 and decrease NDI. In the patients with severe disability, GAMER-MRIL identified smaller relevant regions within MS lesions and larger relevant regions in the corticospinal tract and collateral fibers to the tract [33], which underlies the role of lesion tissue and of damage to main motor tract in disability worsening. Future works should aim to apply this method to a larger cohort of MS patients to further corroborate these findings.

Our approach is novel. Recent works demonstrated the applicability of LRP on MS qualitative MRI datasets by deploying CNN and LRP in the classification related to MS patients. These works used single rules and their relevance maps based on the classification probability instead of the logits. In [10], [34], healthy controls and MS patients were classified using single-path CNNs, and then the authors applied the ε-rule with different ε values to obtain their relevance maps for analyses. In [11], the authors trained single-path CNNs to classify relapsing-remitting MS (RRMS) and progressive MS (PMS) patients, and the αβ-rule with α=1, β=0 was used. In [12], they used a single-path CNN to classify MS patients with EDSS≥3 and applied the αβ-rule with α=1, β=0. The obtained LRP maps showed high relevance surrounding and on the edges of the ventricles, which reflected more on the presence of lesions instead of the symptom-related brain areas. Different from those previous works, GAMER-MRIL is a multi-path attention-based CNN to simultaneously consider the information of tissue properties provided by multiple qMRIs and the obtained relevance map is more relevant to the symptom, i.e. mobility impairment in this study. Overall, GAMER-MRIL provides a unique way to identify regions within the brain whose alteration in qMRI maps is strongly associated with clinical features in MS patients.

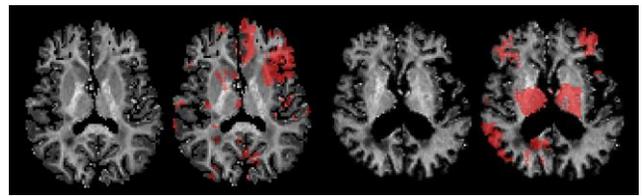

Fig. 5. The top $40^{th}$ quantile masks of the scenario 3 on two exemplary patients' NDI images. Leftmost: the NDI image of a patient in the mild disability group; Middle left: the leftmost NDI image overlapped with the mask; Middle right: the NDI image of a patient in the severe disability; Rightmost: the middle right NDI image overlapped with the mask

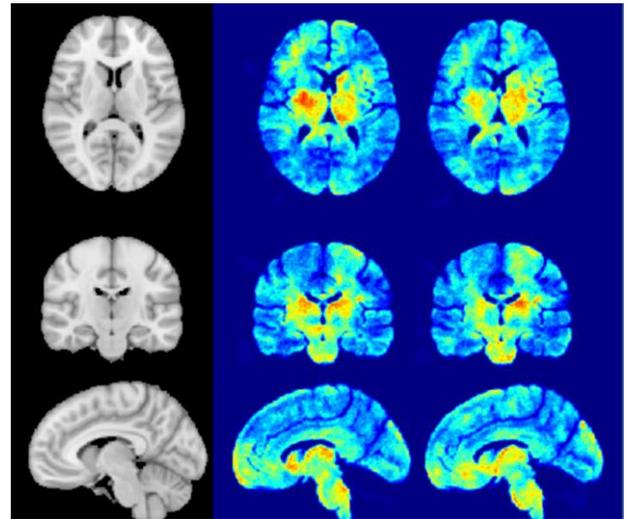

Fig. 6. The heatmaps of the top $40^{th}$ quantile masks of the scenarios 3 and 4 on the MNI152 template. Left: MNI152 template Middle: the heatmap of the scenario 3. Right the heatmap of the scenario 4. The color scale is from 0 to 0.65.

## V. Limitation

There were some limitations to this work. The first limitation was the size of the dataset. Even though the size of the dataset is large compared with other neuroimaging studies exploiting qMRI and DenseNet greatly reduces the number of learnable parameters, the original DenseNet configuration with the parallel encoding structure of the network in GAMER-MRIL still led to a sizeable number of parameters. It was easily prone to overfitting using qMR images in this work. On the hand, the richness of hidden representation of the DenseNet in this work was limited to certain degrees due to this minimal DenseNet configuration. Transfer learning can alleviate this issue and was demonstrated in [34] using the Alzheimer's Disease Neuroimaging Initiative dataset. We have attempted to pre-train the model on the same patients' qualitative MRI contrasts and fine-tuned it on the qMR images, but the model performance did not improve. It might be that the representation of qualitative MRI contrasts learned by the pre-trained model was quite different from the target application using the qMR images, and the model suffered from the negative transfer issue [35]. The second limitation was the choice of LRP rules. Here we only utilized rules including the ε-rule, the αβ-rule, and the $z^\beta$-rule in addition to the proposed and LRP-all rules. There are more LRP rules for different types and positions of layers [36], [37], and it might be beneficial for a more comprehensive assessment to experiment on the best combinations of rules. Another limitation was that the relevance maps and the combined map were used as different quantile masks instead of the values being used. The relevance value received by a voxel was often tiny due to numerous voxels. Results in [34] and [11] were also affected by the tiny values. The value itself could be affected differently across data samples depending on the choice of ε in the ε-rule and the numerical precision used for training the NN. In [34], ε was 0.001, whereas it was 1 in [10]. The use of the mask based on quantiles of the relevance values functions as a workaround to this issue. Furthermore, only the positive relevance values were considered, and the assessment of the negative relevance values will be a potential future work.

## VI. Conclusion

In summary, we demonstrate that GAMER-MRIL can classify patients of severe disability using volumetric qMR images with extended GAMER-MRI and subsequently identify brain regions that are most related to patients' disability with the new LRP approach.

Future work will aim at (i) investigating the hidden representations learned by GAMER-MRIL and their pathological meaning; (ii) integrating other qMRI measures such as quantitative susceptibility mapping (QSM) and magnetization transfer saturation (MTsat) and (iii) increasing the richness and completeness of the representations by the incorporation of bias parameters during LRP backward pass.

## Acknowledgment

This work was supported by Swiss National Funds PZ00P3 154508, PZ00P3 131914 and PP00P3 176984. We would like to acknowledge all the patients and healthy controls in this project.